\documentclass{emulateapj}

\shorttitle{OGLE-III Cepheid P-L Relations}
\shortauthors{Ngeow et al.}

\begin{document}

\title{Period-Luminosity Relations Derived from the OGLE-III Fundamental Mode Cepheids}

\author{Chow-Choong Ngeow}
\affil{Department of Astronomy, University of Illinois, Urbana-Champaign, IL 61801}

\author{Shashi M. Kanbur}
\affil{Department of Physics, State University of New York at Oswego, Oswego, NY 13126} 

\author{Hilding R. Neilson}
\affil{Department of Astronomy \& Astrophysics, University of Toronto, Toronto, ON, Canada M5S 3H4} 

\author{A. Nanthakumar}
\affil{Department of Mathematics, State University of New York at Oswego, Oswego, NY 13126}

\and

\author{John Buonaccorsi}
\affil{Department of Mathematics \& Statistics, University of Massachusetts, Amherst, MA 01003}

\begin{abstract}

In this Paper, we have derived Cepheid period-luminosity (P-L) relations for the Large Magellanic Cloud (LMC) fundamental mode Cepheids, based on the data released from OGLE-III. We have applied an extinction map to correct for the extinction of these Cepheids. In addition to the $VIW$ band P-L relations, we also include $JHK$ and four {\it Spitzer IRAC} band P-L relations, derived by matching the OGLE-III Cepheids to the 2MASS and SAGE datasets, respectively. We also test the non-linearity of the Cepheid P-L relations based on extinction-corrected data. Our results (again) show that the LMC P-L relations are non-linear in $VIJH$ bands and linear in $KW$ and the four {\it IRAC} bands, respectively.

\end{abstract}


\section{Introduction}

Recently, the Optical Gravitational Lensing Experiment (OGLE) team released a catalog of the Large Magellanic Cloud (LMC) Cepheids from its third phase of observation (hereafter OGLE-III), as described in \citet{sos08}. In their paper, the Cepheid period-luminosity (P-L) relations in the $V$ and $I$ bands were derived from this OGLE-III catalog without any extinction correction, and it was left for other researchers to correct for the extinction. On the other hand, the current optical band LMC P-L relations available in the literature \citep[see, for example,][]{uda99a,san04,kan03,kan06,fou07} are mostly derived from the OGLE-III predecessor, the OGLE-II catalog \citep{uda99b}. In the $JHK$ band, \citet{per04} derived the P-L relation based on the complete light curves of 92 LMC Cepheids, while \citet{gro00} and \citet{fou07} derived the $JHK$ P-L relations by matching the 2MASS (Two-Micron All Sky Survey) point sources to the OGLE-II Cepheids. At longer wavelengths, the {\it Spitzer IRAC} band P-L relations were derived by matching the OGLE-II Cepheids and the Cepheids sample from \citet{per04}, done in \citet{nge08b} and \citet{fre08} respectively, to the SAGE (Surveying the Agents of a Galaxy's Evolution) archival data.  

In this Paper, we attempt to derive the extinction corrected $VI$ band P-L relations by applying an extinction map to the OGLE-III Cepheids \citep[which is not done in][]{sos08}. We also derive the $JHK$ and the {\it IRAC} band P-L relations by matching the OGLE-III Cepheids to the 2MASS and SAGE database, respectively. In addition, one unexpected result from the OGLE-II based P-L relation is that the optical LMC P-L relation was found to be non-linear \citep[see, for example,][and reference therein]{san04,kan04,nge08a}. We will also test the non-linearity of the extinction corrected OGLE-III P-L relations using rigorous statistical tests. 

\section{The Data}

The data for OGLE-III fundamental mode (FU) Cepheids were kindly provided for us by I. Soszy\'{n}ski: this included the celestial coordinates, the mean $VI$ band magnitudes, and the periods for $1848$ Cepheids. Further details about this dataset can be found in \citet{sos08}. Throughout this Paper, we have assumed the classification of these FU Cepheids, as given in \citet{sos08}, is robust. Five Cepheids without $V$ and $I$ band mean magnitudes were excluded from the sample. We also excluded the 11 Cepheids with $\log P>1.5$ (where $P$ is the pulsation period) that only have $V$ band data. The observed magnitudes of these Cepheids are close to the saturation limit of the OGLE-III survey \citep[at $I\sim13$ mag.,][]{sos08}, hence their photometry may be affected and/or biased. 

We also matched the OGLE-III FU Cepheids with the 2MASS point source catalog \citep{cut03,skr06}, using a search radius of $1$'', in order to derive the corresponding P-L relations in the $JHK$ bands. The match did not find the 2MASS counterparts for $71$ out of the $1832$ Cepheids in our sample within the search radius. The remaining $1761$ Cepheids only have one matched 2MASS point source with a mean separation of $0.108$'' (RMS $=0.110$''). We applied the prescription outlined in \citet{sos05} to convert the single-epoch 2MASS photometry to the mean $JHK$ magnitudes, using the epoch of the 2MASS data, the period and the epoch at maximum light of the Cepheids, and the scaling between the $I$ band and $JHK$ band amplitudes.

Finally, we matched the OGLE-III FU Cepheids with the SAGE data \citep{mei06} from {\it Spitzer} observation. The {\it IRAC} band data was taken from the SAGE Winter '08 IRAC Archive\footnote{Using the IRSA's Gator Catalog Query at {\tt http://irsa.ipac.caltech.edu/applications/Gator/}}. In contrast to the SAGE data used in \citet{nge08b}, this new SAGE data contains the photometry from both Epoch 1 and 2 observations. Using the same $1$'' search radius, the query to the SAGE Archive returned $1781$ and $1774$ unique sources for the Epoch 1 and 2, respectively, and $1759$ of them have both the Epoch 1 and 2 photometry. The mean separations for the matched Epoch 1 and 2 sources are $0.222$'' (RMS $=0.114$'') and $0.163$'' (RMS $=0.147$''), respectively. There are a total of $1796$ sources after combining the data from the two epochs. We took the averaged values from the two epochs for the matched sources, or kept the single epoch magnitudes if either one of the epochs is available.

\section{The Extinction Corrected P-L Relations}

There are various ways to estimate the extinction for Cepheids \citep[for example, see the introduction in][]{kov08}. In this Paper, we adopted the extinction map from \citet[][hereafter Z04]{zar04} as a mean to estimate the extinction for individual OGLE-III FU Cepheids, because the Z04 extinction map covers almost the entire LMC. The extinction map from \citet{ima07} covers a similar region as in Z04 extinction map, but their extinction map is not publicly available. The OGLE team has also published the LMC extinction map based on the OGLE-II red-clump stars \citep{uda99b}, which was later refined by \citet{sub05}, however the OGLE-II extinction map only covers the central bar region in the LMC. Comparison of the different extinction maps and various methods is beyond the scope of this Paper, and will be presented in future papers. As in \citet{nge05}, we selected the extinction values derived from the cool stars when applying the Z04 extinction map. For a given Cepheid location, the map returns the $V$ band extinction $A_V$. For the three Cepheids that returned a null value of $A_V$, we used the mean extinction value of $E(B-V)=0.10$ appropriate for the LMC. The averaged $A_V$ for the OGLE-III Cepheids is $0.429$, which translates to a mean $E(B-V)$ of $0.129$. For extinction in other bands, we scaled the $A_V$ values with $R_{\lambda}/R_V$, where $R$ is the total-to-selective extinction coefficient. The adopted values for $R$ are $R_{V,I,J,H,K,3.6,4.5,5.8,8.0}=\{3.24,\ 1.96,\ 0.95,\ 0.59,\ 0.39,\ 0.17,\ 0.12,\ 0.08,\ 0.05\}$. The values in $VI$ band are taken from \citet{uda99b}, while for other bands, the values are derived from the \citet{car89} extinction law. In addition, we also derived the reddening-free Wesenheit function $W=I-1.55(V-I)$, with the same expression given in \citet{uda99a}, in this Paper.

\begin{figure}
\plotone{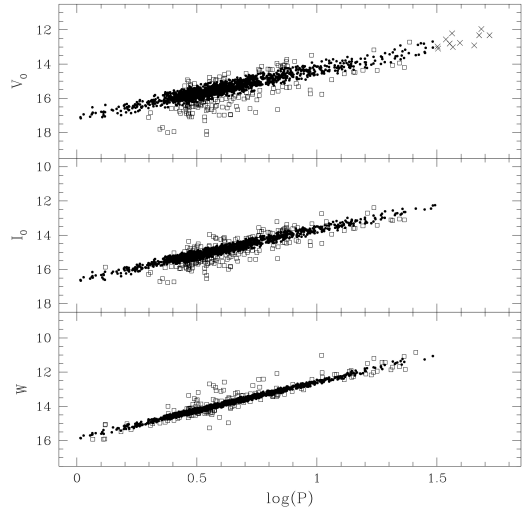}
\caption{The $V$, $I$ and $W$ band P-L relations derived from the OGLE-III fundamental mode Cepheids. The $V$ and $I$ band mean magnitudes have been corrected for extinction using the extinction map from Z04. The filled circles are the Cepheids used to derive the P-L relations, and the open squares are the rejected outliers. The crosses are the excluded 11 long period Cepheids as described in the text.\label{OGLE_PL}}
\end{figure}

\begin{figure}
\plotone{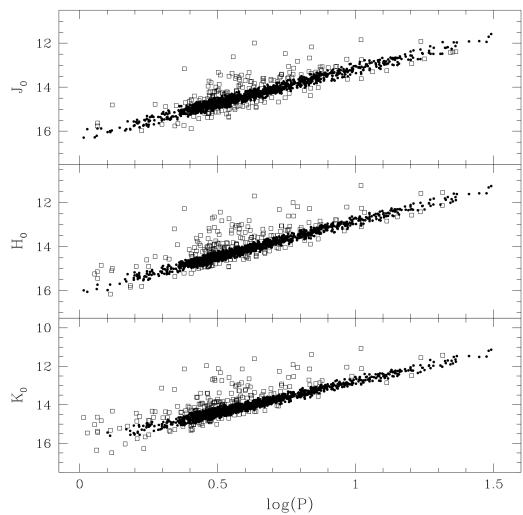}
\caption{The $J$, $H$ and $K$ band P-L relations derived from matching the OGLE-III fundamental mode Cepheids with 2MASS point source catalog. The $JHK$ band magnitudes have been corrected for random phase and extinction as described in the text. The symbols are same as in Figure \ref{OGLE_PL}. \label{2MASS_PL}}
\end{figure}

\begin{figure}
\epsscale{.9}
\plotone{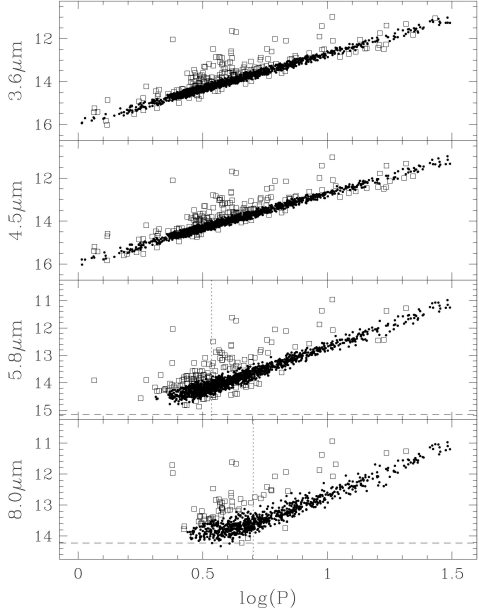}
\caption{The {\it IRAC} band P-L relations derived from matching the OGLE-III fundamental mode Cepheids with SAGE Archival data. The symbols are same as in Figure \ref{OGLE_PL}. The dashed and dotted lines on the lower two panels are the detection limit and the adopted period cut in those bands, respectively. \label{SAGE_PL}}
\end{figure}

\begin{deluxetable}{lccc}
\tabletypesize{\scriptsize}
\tablecaption{The LMC P-L Relation in Various Bands.\label{tabpl}}
\tablewidth{0pt}
\tablehead{
\colhead{Band} &
\colhead{Slope} &
\colhead{Zero-Point} &
\colhead{$\sigma$} 
}
\startdata
$V$      & $-2.769\pm0.023$ & $17.115\pm0.015$ & 0.204  \\
$I$      & $-2.961\pm0.015$ & $16.629\pm0.010$ & 0.131  \\
$J$      & $-3.115\pm0.014$ & $16.293\pm0.009$ & 0.121  \\
$H$      & $-3.206\pm0.013$ & $16.063\pm0.008$ & 0.107  \\
$K$      & $-3.194\pm0.015$ & $15.996\pm0.010$ & 0.124  \\
$3.6\mu$m& $-3.253\pm0.010$ & $15.967\pm0.006$ & 0.087  \\
$4.5\mu$m& $-3.214\pm0.010$ & $15.930\pm0.006$ & 0.089  \\
$5.8\mu$m& $-3.182\pm0.020$ & $15.873\pm0.015$ & 0.122  \\
$8.0\mu$m& $-3.197\pm0.036$ & $15.879\pm0.034$ & 0.139  \\
$W$      & $-3.313\pm0.008$ & $15.892\pm0.005$ & 0.069  
\enddata
\end{deluxetable}
 
Outliers of the P-L relations were removed using an iterative $2.5\sigma$ clipping algorithm \citep[for example, see][]{uda99a}, where $\sigma$ is the RMS derived from fitting the P-L relation to the data in a given band. The resulting P-L relations are presented in Figure \ref{OGLE_PL}, \ref{2MASS_PL} and \ref{SAGE_PL}, respectively. For the $5.8\mu\mathrm{m}$ and $8.0\mu\mathrm{m}$ P-L relations, we an applied additional period cut at the short period end. This is because the magnitudes for the short period Cepheids are approaching the detection limits at $\sim15.2$mag. and $\sim14.2$mag. in the $5.8\mu\mathrm{m}$ and $8.0\mu\mathrm{m}$ band \citep{mei06}, respectively. Further, the $8.0\mu\mathrm{m}$ P-L relation displays a flattening for Cepheids with $\log P<0.7$, which is probably due to the larger photometric errors when approaching the limiting magnitudes. A period cut is needed to remove the bias due to this effect. We adopt a period cut at $\log P_{\mathrm{cut}}\sim0.536$ and $\sim0.702$ for the $5.8\mu\mathrm{m}$ and $8.0\mu\mathrm{m}$ P-L relations, respectively. These period cuts were estimated from a non-linear parametric estimation procedure (same procedure to estimate the break period in the next section). The slopes become shallower without these period cuts. The {\it linear} version of the P-L relations in all bands is summarized in Table \ref{tabpl}. Note that we obtained an (almost) identical $W$ band P-L relation as given in \citet{sos08}.

\subsection{Comparison with the Published Results}

In Table \ref{tabcompare}, we compare our multi-band P-L relations from Table \ref{tabpl} with the published optical $VIW$ band P-L relations derived from the OGLE-II Cepheids \citep{uda99a,fou07}, as well as the $JHK$ band P-L relations \citep{fou07} and the {\it IRAC} band P-L relations \citep{nge08b} that matched the OGLE-II Cepheids to the 2MASS and SAGE (Epoch 1) data, respectively. In this comparison, our philosophy is that it is not sufficient to simply compare whether the P-L slope and/or zero-point obtained by one sample is within one or two standard deviations of the P-L slope and/or zero-point from another sample: this is a necessary but not sufficient condition. What is needed is a simultaneous comparison. An analytical argument showing this has been stated in \citet{nge08a}. 

\begin{deluxetable}{lccccc}
\tabletypesize{\scriptsize}
\tablecaption{Comparison of Various P-L Relations.\label{tabcompare}}
\tablewidth{0pt}
\tablehead{
\colhead{Ref.} &
\colhead{Slope} &
\colhead{Zero-Point} &
\colhead{$N$} &
\colhead{$|T|$} &
\colhead{$t$} 
}
\startdata
\multicolumn{6}{c}{$V$ Band} \\
1      & $-2.769\pm0.023$ & $17.115\pm0.015$ & 1675 & $\cdots$ & $\cdots$ \\
2      & $-2.760\pm0.031$ & $17.042\pm0.021$ & 649  & 0.208 & 0.836 \\ 
3      & $-2.779\pm0.031$ & $17.066\pm0.021$ & 650  & 0.232 & 0.817 \\
4      & $-2.734\pm0.029$ & $17.052\pm0.007$ & 716  & 0.938 & 0.348 \\
\multicolumn{6}{c}{$I$ Band} \\
1      & $-2.961\pm0.015$ & $16.629\pm0.010$ & 1640 & $\cdots$ & $\cdots$ \\
2      & $-2.962\pm0.021$ & $16.558\pm0.014$ & 658  & 0.036 & 0.972 \\
3      & $-2.979\pm0.021$ & $16.594\pm0.014$ & 662  & 0.636 & 0.525 \\	
4      & $-2.957\pm0.020$ & $16.589\pm0.005$ & 692  & 0.160 & 0.873 \\
\multicolumn{6}{c}{$W$ Band} \\
1      & $-3.313\pm0.008$ & $15.892\pm0.005$ & 1501 & $\cdots$ & $\cdots$ \\
2      & $-3.277\pm0.014$ & $15.815\pm0.010$ & 690  & 2.321 & 0.020 \\
3      & $-3.309\pm0.011$ & $15.875\pm0.007$ & 671  & 0.269 & 0.788 \\
4      & $-3.320\pm0.011$ & $15.880\pm0.003$ & 686  & 0.517 & 0.605 \\
\multicolumn{6}{c}{$J$ Band} \\
1      & $-3.115\pm0.014$ & $16.293\pm0.009$ & 1586 & $\cdots$ & $\cdots$ \\
4      & $-3.139\pm0.026$ & $16.273\pm0.006$ & 529  & 0.836 & 0.403 \\
\multicolumn{6}{c}{$H$ Band} \\
1      & $-3.206\pm0.013$ & $16.063\pm0.008$ & 1561 & $\cdots$ & $\cdots$ \\
4      & $-3.237\pm0.024$ & $16.052\pm0.005$ & 529  & 1.186 & 0.236 \\
\multicolumn{6}{c}{$K$ Band} \\
1      & $-3.194\pm0.015$ & $15.996\pm0.010$ & 1554 & $\cdots$ & $\cdots$ \\
4      & $-3.228\pm0.028$ & $15.989\pm0.006$ & 529  & 1.120 & 0.263 \\
\multicolumn{6}{c}{$3.6\mu$m Band} \\
1      & $-3.253\pm0.010$ & $15.967\pm0.006$ & 1617 & $\cdots$ & $\cdots$ \\
5      & $-3.263\pm0.016$ & $15.945\pm0.012$ & 628  & 0.566 & 0.572 \\
\multicolumn{6}{c}{$4.5\mu$m Band} \\
1      & $-3.214\pm0.010$ & $15.930\pm0.006$ & 1633 & $\cdots$ & $\cdots$ \\
5      & $-3.221\pm0.017$ & $15.927\pm0.012$ & 635  & 0.377 & 0.707 \\
\multicolumn{6}{c}{$5.8\mu$m Band} \\
1      & $-3.182\pm0.020$ & $15.873\pm0.015$ & 931  & $\cdots$ & $\cdots$ \\
5      & $-3.173\pm0.028$ & $15.850\pm0.022$ & 561  & 0.272 & 0.785 \\
\multicolumn{6}{c}{$8.0\mu$m Band} \\
1      & $-3.197\pm0.036$ & $15.879\pm0.034$ & 401  & $\cdots$ & $\cdots$ \\
5      & $-3.091\pm0.039$ & $15.684\pm0.036$ & 319  & 1.954 & 0.051 
\enddata
\tablecomments{Reference: (1) this work; (2) \citet{uda99a}; (3) same as (2) but with the updated version given in {\tt ftp://sirius.astrouw.edu.pl/ogle/ogle2/var\_stars/lmc/cep/catalog/\\ README.PL}; (4) \citet{fou07}; (5) \citet{nge08b}. }
\end{deluxetable}

We applied the standard $t$-test \citep[for example, see][]{bet95} to test the published P-L slopes in Table \ref{tabcompare}, under the null hypothesis that the slope derived in this Paper, for a given band, is consistent with the published slopes. The $T$-values calculated from the $t$-test incorporate the standard deviations of the slopes and the sample variance in the two samples under testing. The corresponding $t_{(\alpha/2,\nu)}$-values were then evaluated from the $t$ distribution with $\nu=N_1+N_2-4$ degree of freedom, where $N_1$ and $N_2$ are the number of Cepheids in the two samples. In our test, we adopted a constant significance level of $\alpha=0.05$. The values of $T$ and $t$ for each pair of the slopes were given in the last two columns of Table \ref{tabcompare}. The null hypothesis of equal slopes can be rejected if $|T|>t$. 

In general, both agreements and disagreements were found when comparing the slopes to the published results. The disagreements include the $VJHK$ band slopes when compared to those published in \citet{fou07}, the $I$ band slope when compared to the updated version of \citet{uda99a}, the $W$ band slope when compared to the slope given in \citet{uda99a}, and the $8.0\mu\mathrm{m}$ band slope when compared it to the slope found in \citet{nge08b}. However, agreements were also found for other $VIW$ slopes in the same band. In all cases, the slopes that are in disagreement are either shallower (in $VW$ band) or steeper (in $I$ band) than the rest of the slopes in the same bands. In the $JHK$ band, the slopes from Table \ref{tabpl} were shallower than the counterparts given in \citet{fou07}. Since both sets of P-L relations matched the LMC Cepheids to the 2MASS data and applied the same transformation of the single epoch photometry to the mean magnitudes \citep[as given in][]{sos05}, the discrepancy of the slopes is due to the different sample size used in both studies \citep[there are about three times more Cepheids in our samples than those in][]{fou07}\footnote{If restricted to $\sim610$ OGLE-III Cepheids that are common to OGLE-II, the P-L slopes we obtained in $JHK$ band are $-3.135\pm0.022$, $-3.228\pm0.020$ and $-3.213\pm0.024$, respectively. When compared to the P-L relations from \citet{fou07}, the corresponding $T$- and $t$-values are $|T_{J,H,K}|=\{0.118,0.290,0.409\}$ and $t_{J,H,K}=\{0.906,0.772,0.683\}$.}. The slopes in $3.6\mu\mathrm{m}$, $4.5\mu\mathrm{m}$ and $5.8\mu\mathrm{m}$ P-L relations agree with those given in \citet{nge08b}, though the result for $3.6\mu\mathrm{m}$ band is marginal. The discrepancy for the $8.0\mu\mathrm{m}$ band P-L relation is mainly due to the additional period cut applied in this Paper. 

The difference of the zero-points in the $VI$ band P-L relations between Table \ref{tabpl} and the published results is mainly due to the different extinction maps used in deriving the P-L relations. The P-L relations given in \citet{uda99a} and \citet{fou07} were derived using the OGLE-II extinction map, with a mean $E(B-V)=0.147$ \citep{uda99a}. This value is higher than the mean $E(B-V)=0.129$ found in this paper from the Z04 extinction map \citep[see a similar finding, for example, in][]{sub05,fin07}. Hence the derived P-L zero-points will be fainter by $\sim0.06$mag. and $\sim0.03$mag. in the $V$ and $I$ band, respectively, to those OGLE-II based P-L relations. This difference is indeed seen in Table \ref{tabcompare} \citep[except for the $I$ band P-L zero-point from][]{uda99a}. The difference of the P-L zero-points in other bands, that less affected by extinction, may be due to the different sample size used in deriving the P-L relations \citep[see also][]{sos08}. 

\section{Testing for Non-Linear P-L Relations}

As in previous work, the non-linearity of the P-L relation is referred to the two P-L relations separated at 10 days \citep{san04,kan04,nge05,nge08a}. We have applied a non-linear parametric estimation procedure available in the {\tt SAS} package, which utilizes the Levenberg - Marquardt algorithm, to estimate the break period from the $I$ band data (because the $I$ band light curves are well sampled from large number of observations). The estimated break period is located at $\log P_{\mathrm{break}}=1.006$ (with the 95\% confidence limits given by $\pm0.170$), which is fully consistent to the adopted 10 day period in defining the non-linearity of the P-L relation.

The fitted long ($\log P>1.0$) and short period P-L relations are summarized in Table \ref{tabf}. At first glance, the difference of the slopes between the long and short period P-L relations in $V$, $K$, $3.6\mu\mathrm{m}$, $4.5\mu\mathrm{m}$, $5.8\mu\mathrm{m}$ and $W$ band seems to be within $\sim1.5\sigma$. This may suggest these P-L relations are linear. In contrast, the $I$, $J$, $H$ and $8.0\mu\mathrm{m}$ band P-L relations show a difference of the slopes at $\sim2.0\sigma$ level or larger. We emphasize again that {\it testing} for non-linearity by simply seeing if the long and short period slopes are within a certain number of standard deviations of each other is a necessary but not a sufficient condition: a simultaneous comparison using statistical test is needed. Thus a P-L relation can be non-linear even if the short and long period slopes are within $1$ to $\sim2\sigma$ of each other as analytically outlined in \citet{nge08a}. Furthermore, \citet{nge06} also argued that statistical tests are needed to detect the non-linear P-L relation.

\begin{deluxetable*}{lcccccccccc}
\tabletypesize{\scriptsize}
\tablecaption{$F$-Test Results of the P-L Relations.\label{tabf}}
\tablewidth{0pt}
\tablehead{
\colhead{Band} &
\colhead{Slope$_S$} &
\colhead{Zero-Point$_S$} &
\colhead{$\sigma_S$} &
\colhead{$N_S$} &
\colhead{Slope$_L$} &
\colhead{Zero-Point$_L$} &
\colhead{$\sigma_L$} &
\colhead{$N_L$} &
\colhead{$F$} &
\colhead{$p(F)$} 
}
\startdata
$V$        & $-2.823\pm0.031$ & $17.143\pm0.018$ & 0.202 & 1566 & $-2.746\pm0.165$ & $17.122\pm0.195$ & 0.230 & 109 & 3.18 & 0.042 \\
$I$        & $-3.004\pm0.020$ & $16.651\pm0.012$ & 0.130 & 1553 & $-2.775\pm0.111$ & $16.440\pm0.129$ & 0.132 &  96 & 5.77 & 0.003 \\
$J$        & $-3.150\pm0.019$ & $16.312\pm0.011$ & 0.119 & 1486 & $-2.909\pm0.120$ & $16.075\pm0.139$ & 0.142 & 100 & 5.24 & 0.005 \\
$H$        & $-3.246\pm0.017$ & $16.085\pm0.010$ & 0.106 & 1461 & $-2.989\pm0.096$ & $15.832\pm0.113$ & 0.117 & 100 & 7.90 & 0.000 \\
$K$        & $-3.212\pm0.021$ & $16.006\pm0.013$ & 0.124 & 1449 & $-3.057\pm0.101$ & $15.845\pm0.117$ & 0.125 & 105 & 1.57 & 0.209 \\
$3.6\mu$m  & $-3.254\pm0.013$ & $15.967\pm0.008$ & 0.086 & 1509 & $-3.233\pm0.068$ & $15.944\pm0.082$ & 0.100 & 108 & 0.06 & 0.946 \\
$4.5\mu$m  & $-3.228\pm0.014$ & $15.938\pm0.008$ & 0.088 & 1524 & $-3.185\pm0.071$ & $15.903\pm0.085$ & 0.104 & 109 & 1.05 & 0.352 \\
$5.8\mu$m  & $-3.164\pm0.035$ & $15.861\pm0.025$ & 0.122 &  813 & $-3.275\pm0.083$ & $15.984\pm0.099$ & 0.125 & 118 & 0.78 & 0.460 \\
$8.0\mu$m  & $-3.036\pm0.100$ & $15.746\pm0.083$ & 0.142 &  286 & $-3.348\pm0.086$ & $16.058\pm0.102$ & 0.127 & 115 & 2.70 & 0.069 \\
$W$        & $-3.329\pm0.010$ & $15.900\pm0.006$ & 0.068 & 1585 & $-3.338\pm0.084$ & $15.934\pm0.096$ & 0.084 &  84 & 2.90 & 0.055
\enddata
\tablecomments{$\sigma$ is the dispersion of the P-L relation. Subscripts $_S$ and $_L$ refer to the short ($\log P<1.0$) and long period Cepheids, respectively. The $p(F)$ is the probability under the null hypothesis of single line regression, for the given $F$ value and the degree of freedom.} 
\end{deluxetable*}

In this Paper, we only use the $F$-test, though several other statistical tests have been applied to the OGLE-II based P-L relations before \citep{kan07,koe07,nge08a}. Details for the $F$-test can be found in \citet{kan04} and \citet{nge05}, and will not be repeated here. For a large number of data points, $F\sim3.0$ at 95\% confidence level. Hence, the underlying P-L relation is statistically consistent  with non-linearity if $F>3$ (that is, the null hypothesis of linear regression can be rejected). The choice of a 95\% confidence level is fairly standard and consistent with our previous work. 

The robustness of the $F$-test has a large literature. Its assumptions are homoskedasticity, normality of errors and independent observations. The last assumption regarding independent observations means that observations of one Cepheid have no bearing on subsequent observations of another Cepheid. \citet{kan04} checked assumptions of normality and homoskedasticity using OGLE-II Cepheids data and found them to be reasonable assumptions. Homoskedasticity essentially requires there to be not too many outliers in the data, which have been removed in our samples. \citet{ali96} suggest the $F$-test is robust to small departures from normality and our large sample size, both with the OGLE-II and OGLE-III Cepheids, further adds to this robustness. For example, \citet{nge06} appended data from different sources to increase the OGLE-II sample size and found the $F$-test produced consistent results.

The $F$-test results were given in the last two columns of Table \ref{tabf}. From this table, The $F$-test results clearly suggest that the $IJH$ band P-L relations are non-linear, in contrast the $K$ band P-L relation and the $3.6\mu\mathrm{m}$ to $5.8\mu\mathrm{m}$ P-L relations are linear. These results support the previous finding of the non-linear/linear LMC P-L relations from the OGLE-II and MACHO Cepheids \citep{nge05,kan06,nge08a,nge08b}, and extend the non-linear/linear results further to the Cepheids from the OGLE-III sample. The $F$-test results on the $V$, $W$ and $8.0\mu\mathrm{m}$ P-L relations are worth a further discussion.

In our previous work, the LMC $V$ band P-L relation shows a highly significant $F$-test result for non-linearity \citep{nge05,kan06,nge08a}, with a large $F$-value and a $p$-value that is significantly smaller than $0.05$. The $F$-value listed in Table \ref{tabf} is $3.18$, which appears to be close to the $F=3.0$ threshold at $95$\% confidence level\footnote{If the confidence level is lowered, say to $90$\%, then the corresponding $F$-value will be decreased to $\sim2.3$. In this case the $F$-test result for the $V$ band P-L relation clearly shows that it is non-linear.} However, hypothesis testing works with $p$-values for the chosen statistics under the null hypothesis. Hence, the probability of having an $F$-value as large as $3.18$ is smaller than $0.05$. Therefore, the $V$ band P-L relation {\it does} show evidence of non-linearity from the $F$-test. 

For the Wesenheit function, \citet{nge05w} and \citet{nge08a} found that the LMC Wesenheit function, using the OGLE-II data, derived from the optical band is linear. The $F=2.90$ (with $p=0.055$) from Table \ref{tabf} suggests the $W$ band P-L relation found in this paper is also linear. However, if the $p=0.042$ for the computed $V$ band $F$ statistics in Table \ref{tabf} is considered to be non-linear, then the same attention must be given to the fact that the $p=0.055$ for $W$ band P-L relation which is close to being considered linear. Our explanation for this has been that the non-linearities in the P-L and period-color (P-C) relations cancel out in producing a linear Wesenheit function \citep{nge05w,koe07}. \citet{koe07} also suggested that with a larger sample and more accurate data, the ``cancellation'' of non-linearities in the PL/PC relation will not necessarily be exact leading to a non-linear Wesenheit function. 

\citet{nge08b} found the $8.0\mu\mathrm{m}$ P-L relation is non-linear, in contrast to the linear $F$-test result given in Table \ref{tabf}. This discrepancy is due to the period cut applied at a rather large period \citep[note that no period cut was applied in][]{nge08b}. In fact, the $8.0\mu\mathrm{m}$ P-L relation becomes non-linear if $\log P_{\mathrm{cut}}<0.69$ is adopted. The non-linear $8.0\mu\mathrm{m}$ P-L relation found in \citet{nge08b} could be caused by the incompleteness of the data at the faint end when approaching the detection limit. Another possibility is that mass-loss may plays some role in longer wavelength P-L relations to make the P-L relations non-linear \citep{nei08}. Detailed study of the effect of mass-loss on the {\it IRAC} band P-L relations, based on the OGLE-III Cepheids, is currently underway, and will be presented in a future paper.

\section{Including the Longer Period Cepheids}

In previous sections, we excluded the Cepheids with $\log P>1.5$ from the OGLE-III sample, which only have the $V$ band data. Including these Cepheids the linear version of the $V$ band P-L relation is almost identical to the relation given in Table \ref{tabpl}. However, the slope of the long period P-L relation becomes steeper ($-2.823$) as compared to the slope given in Table \ref{tabf} ($-2.746$) if these Cepheids were included. Previous studies \citep[see, for example,][]{san04,kan06,nge08a} found that the long period $V$ band P-L relation is shallower than the short period counterpart. Hence the steeper slope may suggest these longer period $V$ band Cepheid data should not be included in the sample. 

It is possible to increase the number of long period Cepheids and extend the period coverage beyond $\log P=1.5$, by including the additional Cepheids from the literature. This has been done, for example, in \citet{san04}. However, one criticism has been that the published long period Cepheid data may not be in the same photometric system as in the OGLE survey \citep{fou07}. The on-going ``OGLE Shallow Survey of the LMC Cepheids'' is aimed to observe the longer period LMC Cepheids in the same photometry system as in OGLE-III. Hence, it is better to wait for the data released from this survey before appending additional longer period Cepheids to our sample.

\section{Conclusion}

In this Paper we derive the $VIJHKW$ and the four {\it IRAC} band LMC P-L relations based on the OGLE-III Cepheids. The Cepheid data were corrected for extinction by applying the Z04 extinction map. We matched the OGLE-III Cepheids to the 2MASS point sources, and the single-epoch 2MASS photometry was transformed to the mean magnitudes. For the {\it IRAC} band data, a straight average from the SAGE Epoch 1 and 2 archival data were adopted, which should closer to the mean magnitudes than those used in \citet{nge08b}. 

We compare our P-L relations to the published results using the $t$-test. For the published optical P-L relations that used in the comparison, $2/3$ of them show a good agreement on the P-L slopes to our results. For the $JHK$ and {\it IRAC} band, the $t$-test results show that our P-L slopes disagree and agree (except for the $8.0\mu\mathrm{m}$ band P-L slope) to the published results, respectively. 

We further test the non-linearity of the P-L relations using the $F$-test to the OGLE-III Cepheids, and found that the LMC $VIJH$ band P-L relations are non-linear but linear in $KW$ and the four {\it IRAC} bands, respectively. It has to be cautious that the $5.8\mu\mathrm{m}$ and $8.0\mu\mathrm{m}$ P-L relations may be affected by incomplete bias at the faint end and/or mass-loss. It is worth pointing out that both MACHO and OGLE-III samples show the same non-linear/linear results in $VJHK$ bands, where both samples have relatively large number of Cepheids and cover the similar region in the LMC. 

\acknowledgments

The authors would like to thank I. Soszy\'{n}ski for the help on retrieving the OGLE-III Cepheid data, and an anonymous referee for the suggestions to improve the manuscript. Support for this work was provided by an award issued by JPL/Caltech ({\it Spitzer} grant ID: 50029). SMK acknowledges support from the Chretien International Research Award from the American Astronomical Society. Part of this work is based on archival data obtained with the {\it Spitzer Space Telescope} and the NASA/IPAC Infrared Science Archive, which is operated by the Jet Propulsion Laboratory, California Institute of Technology under a contract with the National Aeronautics and Space Administration. This publication also makes use of data products from the Two Micron All Sky Survey, which is a joint project of the University of Massachusetts and the Infrared Processing and Analysis Center/California Institute of Technology, funded by the National Aeronautics and Space Administration and the National Science Foundation.

\end{document}